\documentclass[
prd,aps,showpacs,tightenlines,nofootinbib,preprintnumbers,
]{revtex4}
\usepackage{graphicx}
\usepackage{amssymb,latexsym}
\usepackage{amsfonts}
\usepackage{amssymb}
\usepackage{cancel}
\usepackage{color}
\input{colordvi.tex}

\newcommand{\bear}{\begin{array}}  \newcommand{\eear}{\end{array}}
\newcommand{\beq}{\begin{equation}}  \newcommand{\eeq}{\end{equation}}
\newcommand{\bef}{\begin{figure}}  \newcommand{\eef}{\end{figure}}
\newcommand{\bec}{\begin{center}}  \newcommand{\eec}{\end{center}}

\newcommand{\beqa}{\begin{eqnarray}}
\newcommand{\eeqa}{\end{eqnarray}}
\newcommand{\p}{\phi}
\newcommand{\simg}{\gtrsim}
\newcommand{\siml}{\lesssim}
\newcommand{\ka}{\kappa}

\newcommand{\Omp}{\Omega_{\phi 0}}

\newcommand{\tlam}{t_{\Lambda}}
\newcommand{\pii}{\phi_i}

\begin{document}

\title{Slow-Roll Thawing Quintessence
}

\author{Takeshi Chiba}%
\affiliation{
Department of Physics, \\
College of Humanities and Sciences, \\
Nihon University, \\
Tokyo 156-8550, Japan}

\date{\today}


\begin{abstract}
We derive slow-roll conditions for thawing quintessence. 
We solve the equation of motion of $\phi$ for a Taylor expanded potential (up to 
the quadratic order) in the limit where the equation of state $w$ is close to $-1$ 
to derive the equation of state as a function of the scale factor.  
We find that the evolution of $\phi$ and hence $w$ are described by only 
two parameters. The expression for $w(a)$, which can be applied to general thawing models,  
coincides precisely with that derived recently by Dutta and Scherrer for hilltop quintessence.  
The consistency conditions of $|w+1|\ll 1$ are derived. The slow-roll conditions 
for freezing quintessence are also derived. 
\end{abstract}

\maketitle

\section{Introduction}

The current cosmological observations seems to be consistent with $\Lambda$CDM. 
The equations of state of dark energy, $w$,  is close to $-1$ within 10\% or less \cite{komatsu}. 
This implies that 
even if a scalar field (dubbed "quintessence") plays the role of dark energy, 
it should roll down its potential slowly because its kinetic energy density should be much 
smaller than its potential. In this situation, as in the case of inflation, 
it is useful to derive the slow-roll conditions for quintessence. 
Quintessence models which hardly move in the past and begin to roll down the potential 
recently are called 
"thawing" models, while "freezing" models move in the opposite ways: they gradually slow down 
the motion \cite{cl}. Our consideration here will be given mostly to thawing models. 

However, as far as we 
are aware, no such conditions have been derived for general potential. 
The usual slow-roll conditions, $(V'/V)^2\ll1$ and $|V''/V|\ll 1$, may not be necessary. 
In fact, in the scalar equation of motion, the "acceleration term", $\ddot\p$, is not 
necessarily small compared with the "friction term", $3H\dot\p$  \cite{ss,cmp,cdl}. 
The difference between inflation and quintessence is that for the former the scalar field always 
dominates the universe, while not for the latter. 
For slow-roll inflation, the time scale of the scalar field motion is longer than the time scale of 
the cosmic expansion which is determined by the potential. 
For quintessence, on the other hand, 
the time scale of the cosmic expansion is determined by the matter/radiation  and 
the scalar field. Therefore the potential 
does not necessarily satisfy the usual slow-roll conditions. 
Note that the situation is not limited to quintessence but is applied to the case when 
the  scalar fields which are subdominant components in 
the universe move slowly. Axion, curvaton, and moduli can be such fields. 

If $w$ is close to $-1$, a functional form of $w$ as a function of the scale 
factor is necessary to parametrize possible deviations from a cosmological constant. 
The frequently used functional form is the linear approximation of $w(a)$ at $a=1$, 
the so-called Chevallier-Polarski-Linder parameterization \cite{cpl}: 
\beqa
w_{\rm linear}(a)=w_0+w_a(1-a).
\label{linder}
\eeqa 
Naively, one may expect that this is a good parameterization because it is a Taylor expansion 
of $w(a)$ at $a=1$. However,  even for a modest redshift $z\simg 2$,   
$1-a$ is not so small $(1-a\simg 0.6)$ and higher order terms $(a-1)^2,(a-1)^3\dots$ may no longer be negligible. 
Recently, it is found that for a particular class of thawing models (hilltop models), 
the equation of motion of the scalar field can be solved explicitly if $|1+w|\ll 1$ 
and $w$ can be written in the closed form which is in excellent agreement with numerical 
solutions \cite{ds}. It is found that even for dark energy with 
$w\simeq -1$, its equation of state cannot evolve like Eq. (\ref{linder}).  
Unfortunately, however, the analysis is limited to the thawing model 
whose potential has a maximum and appears not to be applied to more general thawing models. 
We show that this is not the case and that the analysis can be applied to more general thawing 
models with generic initial conditions. The functional form of $w$ coincides with that derived 
in \cite{ds}. 

In this paper, in Sec. II, we first derive the slow-roll conditions for thawing quintessence models. 
Then,  generalizing the analysis in \cite{ds}, we solve the equation of motion 
of the scalar field for a Taylor expanded potential (up to the quadratic order) in the 
limit where the equation of state $w$ is 
close to $-1$ to derive the equation of state 
$w$ as a function of the scale factor. We also derive the consistency conditions of 
the approximation. 
In Sec. III, we compare the derived $w$ with numerical solutions for several quintessence 
potentials which unlike \cite{ds} do not necessarily have the maximum and find fairly 
good agreement. Sec. IV is devoted to summary. In Appendix \ref{app1}, the slow-roll conditions for freezing quintessence 
are derived, and in Appendix \ref{app2} some calculations which are necessary to derive a 
equation used in the text are given.

\section{Slow-Roll Thawing Quintessence}

Working in units of $8\pi G=1$, the basic equations in a flat universe are
\beqa
&&\ddot\p + 3H\dot\p +V'=0,\label{eomp}\\
&&H^2=\left({\dot a\over a}\right)^2=\frac{1}{3}(\rho_B+\rho_{\p}),\label{eomh}\\
&&\dot H=-\frac12\left((\rho_B+p_B)+(\rho_{\p}+p_{\p})\right)=-\frac12\left((1+w_B)\rho_B+\dot\p^2\right),
\label{eomhdot}
\eeqa
where $V'=dV/d\p$,  $H=\dot a/a$ is the Hubble parameter with $a$ being the scale factor,  
 $\rho_B(p_B)$ is the energy density (pressure) of matter/radiation,  
$\rho_{\p}=\dot\p^2/2+V(\p) (p_{\p}=\dot\p^2/2-V(\p))$ is 
the scalar field energy density (pressure), and $w_B$ is the equation of state of matter/radiation.

\subsection{Slow-Roll Conditions for Thawing Quintessence}

By slow-roll quintessence we mean a model of quintessence whose kinetic 
energy density is much smaller than its potential, 
\beqa
\frac12\dot\p^2\ll V.
\label{slowroll1}
\eeqa
Unlike the case of inflation, we do not require that $\ddot\p$ is smaller than the friction 
term $3H\dot\p$ in Eq. (\ref{eomp}) since $H$ is not determined by the potential alone, but 
by the matter/radiation along with the scalar field energy density. 

With fixed $w_0$, slowly rolling thawing models correspond to the equation of state 
$w=p_{\p}/\rho_{\p}$ very 
close to $-1$, so that the Hubble friction is not effective and hence $\ddot\p$ is not 
necessarily small compared with $3H\dot\p$ in Eq. (\ref{eomp}).  Slowly rolling 
freezing models correspond to models whose $w$ is not so close 
to $-1$ compared with thawing models so that the Hubble friction is effective and 
 $\ddot\p$ is smaller than $3H\dot\p$ in Eq. (\ref{eomp}).

We derive the slow-roll conditions for thawing quintessence during the matter/radiation 
dominated epoch.  For slow-roll conditions for freezing quintessence,  see Appendix \ref{app1}. 
We first introduce the following function \cite{cmp} (see also \cite{linder}):
\beqa
\beta=\frac{\ddot\p}{3H\dot\p}.
\label{beta}
\eeqa
As stated above, for thawing models, $\beta$ is a quantity of  ${\cal O}(1)$. We assume 
$\beta$ is an approximately constant in the sense $|\dot\beta|\ll H|\beta|$, and the consistency of the assumption 
will be checked later. In terms of $\beta$, using Eq. (\ref{eomp}), $\dot\p$ is written as
\beqa
\dot\p =-\frac{V'}{3(1+\beta)H},
\label{pdot}
\eeqa
and the slow-roll condition Eq. (\ref{slowroll1}) becomes
\beqa
\epsilon:=\frac{V'^2}{6H^2V}\ll 1, 
\label{slow:cond:1}
\eeqa
where we have omitted $1+\beta$ since it is an ${\cal O}(1)$ quantity and introduced the factor of 
$1/6$ so that $\epsilon$ coincides with the inflationary slow-roll parameter, 
$\epsilon={1\over 2}(V'/V)^2$ \cite{ll},  if the scalar field dominates the expansion: $H^2\simeq V/3$. 
Eq. (\ref{slow:cond:1}) is a quintessence counterpart of the inflationary slow-roll condition $(V'/V)^2\ll 1$. 

Similar to the case of inflation, the consistency of Eq. (\ref{beta}) and Eq. (\ref{eomp}) 
should give the second slow-roll condition. In fact, from the time derivative of Eq. (\ref{pdot}) 
\beqa
\ddot\p
=\frac{V''V'}{9(1+\beta)^2H^2}-\frac{1+w_B}{2(1+\beta)}V'+\frac{\dot\beta V'}{3(1+\beta)^2H}, 
\eeqa
where we have used $\dot H/H^2\simeq -3(1+w_B)/2$ from Eq. (\ref{eomh}) and Eq. (\ref{eomhdot}). 
On the other hand, from Eq. (\ref{beta}) and Eq. (\ref{pdot}), 
$\ddot\p=3\beta H\dot\p=-\beta V'/(1+\beta)$, and so we obtain 
\beqa
\beta=-\frac{V''}{9(1+\beta )H^2}+\frac{(1+w_B)}{2}-\frac{\dot\beta}{3(1+\beta)H}\simeq 
-\frac{V''}{9(1+\beta) H^2}+\frac{(1+w_B)}{2},
\label{betaeq}
\eeqa
where we have used $\dot\beta\ll H\beta$. While  
the left-hand-side of Eq. (\ref{betaeq}) is an almost time-independent quantity by assumption, 
the first term in the right-hand-side is 
a time-dependent quantity in general. Therefore the equality holds if 
the first term is negligible:
\beqa
\eta:=\frac{V''}{3H^2}; ~~~~~~~~|\eta|\ll 1,
\label{slow:cond:2}
\eeqa
so that $\beta$ becomes
\beqa
\beta=\frac{1+w_B}{2},
\label{betasol}
\eeqa
{\it or} the left-hand-side is negligible: 
\beqa
|\beta|\ll 1,
\label{betacond}
\eeqa
so that
\beqa
\eta= \frac32 (1+w_B).
\label{etasol}
\eeqa
The former condition corresponds to the slow-roll thawing models, while the latter corresponds to 
the slow-roll freezing models (see Appendix \ref{app1}). 
$\beta$ given by Eq. (\ref{betasol}) is an approximately constant, which is  consistent 
with our assumption.\footnote{More precisely,  
$\dot\beta/H\beta=3(c_s^2-w_B)$, where $c_s^2=\dot p_B/\dot\rho_B$ is the adiabatic sound speed. 
Hence the assumption $|\dot\beta|\ll H|\beta|$ is valid if only the adiabatic perturbation is considered. }
Here the factor $1/3$ is introduced in Eq. (\ref{slow:cond:2}) so that $\eta$ coincides with 
the inflationary slow-roll parameter \cite{ll}, 
$\eta=V''/V$,  if $H^2\simeq V/3$. 
Eq. (\ref{slow:cond:2}) is a quintessence counterpart of the inflationary slow-roll condition 
$|V''|/V\ll 1$.

Eq. (\ref{slow:cond:1}) and Eq. (\ref{slow:cond:2}) constitute the slow-roll conditions for thawing 
quintessence during the matter/radiation epoch.\footnote{It is easily found that even if multiple scalar 
fields each with the same potential are introduced, 
the assisted dynamics \cite{assisted} does not occur: $\epsilon$ and $\eta$ remain the same. } 
Moreover once the universe becomes dominated by 
the scalar field, the two conditions reduce to the usual inflationary slow-roll conditions from 
$H^2\simeq V/3$. Therefore, these conditions (Eq. (\ref{slow:cond:1}) and Eq. (\ref{slow:cond:2})) 
are the slow-roll conditions for thawing quintessence at all times, 
both during the matter/radiation era and during the scalar field dominated era.  
Note that since $H^2\simg V/3$, the inflationary slow-roll conditions 
are sufficient conditions for slow-roll thawing quintessence during the matter/radiation era, 
not necessary conditions.  
In Fig. 1, the evolution of $\beta$ is shown for 
a thawing quintessence model ($V=M^4(1-\cos\p))$. The evolution of $\beta$ agrees nicely with 
Eq. (\ref{betasol}). 

\begin{figure}
\includegraphics[width=13cm]{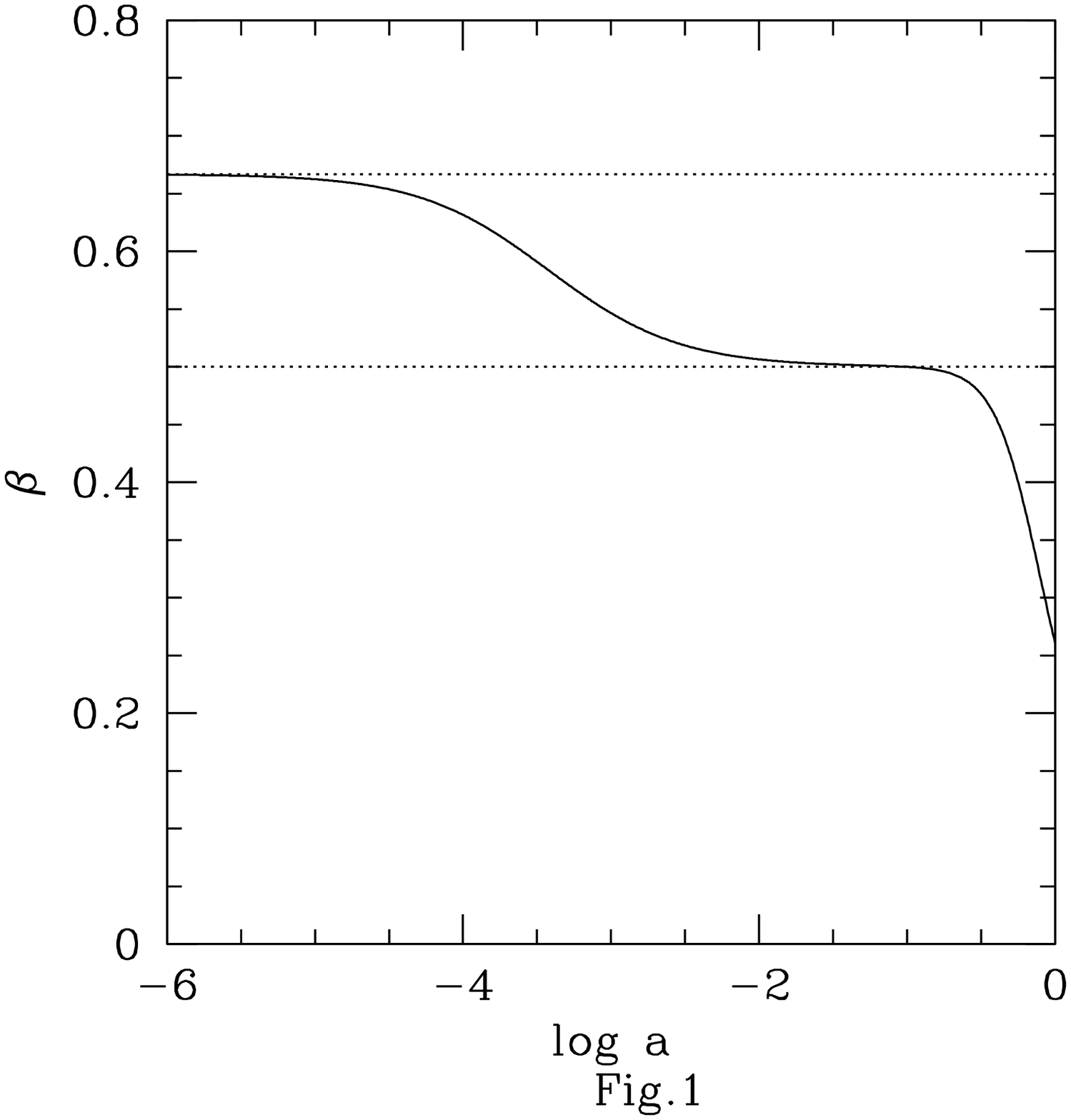}
\caption{ $\beta$ as a function of $a$ for thawing quintessence model with the axion-like potential 
$V=M^4(1-\cos\p)$. The dotted lines are $\beta=2/3, 1/2$, respectively. }
\label{fig1}
\end{figure}

\subsection{Parametrizing the Equation of State}

Next we derive general solutions of $\p$ in the limit of $|1+w|\ll 1$ and derive $w$ as a 
function of $a$. 
To do so, we first note that the Hubble friction term in Eq. (\ref{eomp}) can be eliminated by the 
following change of variable \cite{ds}
\beqa
u=(\p-\p_i) a^{3/2},
\eeqa
where $\p_i$ is an arbitrary constant,  which is introduced for later use, 
and then Eq. (\ref{eomp})  becomes
\beqa
\ddot u+\frac34 (p_B+p_{\p}) u+a^{3/2} V'=0.
\label{eomu}
\eeqa

We assume a universe consisting of matter and quintessence with $w\simeq -1$. 
Then the pressure is well approximated by a constant: $p_B+p_{\p}\simeq p_{\p}\simeq -\rho_{\p 0}$, 
where $\rho_{\p 0}$ is the nearly constant density contributed by the quintessence 
in the limit $w\simeq -1$. Eq. (\ref{eomu}) then becomes
\beqa
\ddot u-\frac34 \rho_{\p 0} u+a^{3/2}V'=0. 
\label{eomu2}
\eeqa
Since we consider a slow-roll scalar field,  the potential may be generally expanded around 
some value $\p_i$, which we identify with the initial value, in the form (up to the quadratic order)
\beqa
V(\p)=V(\p_i)+V'(\p_i)(\p-\p_i)+\frac12V''(\p_i)(\p-\p_i)^2.
\label{exp}
\eeqa
We will check the consistency of the expansion later. 
Substituting the expansion Eq. (\ref{exp}) into Eq. (\ref{eomu2}) and taking 
$\rho_{\p 0}=V(\pii)$  gives
\beqa
\ddot u+\left(V''(\pii)-\frac34 V(\pii)\right)u=-V'(\pii)a^{3/2}.
\label{eomu3}
\eeqa
Here the source term in the right hand side of Eq. (\ref{eomu3}) appears since we consider 
the general Taylor expansion of $V$ around the {\it initial} value $\pii$ in contrast 
with \cite{ds} where the Taylor expansion of $V$ around its {\it maximum} is considered 
and hence $V'$ term in Eq. (\ref{exp}) is absent. 

Being consistent with $|w+1|\ll 1$, we assume $a(t)$ is well approximated by its value in the 
${\rm \Lambda}$CDM model 
which is given by
\beqa
a(t)=\left(\frac{1-\Omp}{\Omp}\right)^{1/3}\sinh ^{2/3}(t/\tlam),
\label{scale}
\eeqa
where $\Omp$ is the present-day value of density parameter of quintessence,  $a=1$ 
at present, and $\tlam$ is defined as
\beqa
\tlam =\frac{2}{\sqrt{3\rho_{\p 0}}}= \frac{2}{\sqrt{3V(\pii)}}.
\eeqa
Introducing 
\beq
k=\sqrt{(3/4)V(\pii)-V''(\pii)},
\eeq
the general solution of Eq. (\ref{eomu3}) is obtained by the use of Green function method 
in the form (if $k\tlam\neq 1$)
\beqa
u(t)=A \sinh(kt)+B\cosh(kt)+
\sqrt{{1-\Omp\over \Omp}}{V'(\pii)\tlam^2\over k^2\tlam^2-1} \sinh(t/\tlam),
\eeqa
where $A$ and $B$ are constants. $k\tlam=1$ corresponds 
to $V''(\pii)=0$, which will be treated separately later. 
As an initial condition, we take that $\p=\pii$ and $\dot\p=\dot\pii$ at $t=t_i$. Then, we obtain
\beqa
\p(t)-\pii=\frac{\sinh(t_i/\tlam)}{k\tlam \sinh(t/\tlam)}\Biggl{[}\sinh(kt)\cosh(kt_i)
\left\{\frac{V'(\pii)}{V''(\pii)}\left(\coth(t_i/\tlam)-k\tlam\tanh(kt_i)\right)+\tlam\dot\pii\right\}\nonumber\\
-\cosh(kt)\sinh(kt_i)\left\{\frac{V'(\pii)}{V''(\pii)}\left(\coth(t_i/\tlam)-k\tlam\coth(kt_i)\right)
+\tlam\dot\pii\right\}\Biggr{]} -\frac{V'(\pii)}{V''(\pii)}.
\label{phi:general}
\eeqa
However, as shown in Appendix \ref{app2}, as long as $a_i\ll 1$ (or $t_i\ll t_0$), 
the solution can be well approximated by that with $t_i=0$
\beqa
\p(t)=\pii+\frac{V'(\pii)}{V''(\pii)}\left( \frac{\sinh(kt)}{k\tlam \sinh(t/\tlam)} -1\right).
\label{phi}
\eeqa

Taking $\rho_{\p}\simeq \rho_{\p 0}\simeq V(\pii)$, using Eq. (\ref{phi}) the equation of state is given by
\beqa
1+w(t)=\frac{\dot\phi^2}{V(\phi_i)}=\frac34 \left(\frac{V'(\pii)}{k\tlam V''(\pii)}\right)^2\left(
\frac{k\tlam \cosh(kt)\sinh(t/\tlam)-
\sinh(kt)\cosh(t/\tlam)}{\sinh^2(t/\tlam)}\right)^2.
\label{eos1}
\eeqa
As done in \cite{ds}, we normalize the expression to the present-day value of $w$, $w_0$, and 
rewrite $w$ as a function of the scale factor using Eq. (\ref{scale}). Normalize 
Eq. (\ref{eos1}) to the present-day value,
\beqa
1+w(a)=(1+w_0)a^{-3}\left(\frac{K\cosh(kt(a))-F(a)\sinh(kt(a))}
{K \cosh(kt_0)-\Omp^{-1/2}\sinh(kt_0)}\right)^2,
\label{eos2}
\eeqa 
where $K=k\tlam$ and $F(a)$ is the inverse square root of the fractional energy density 
corresponding to a cosmological constant and they are given by
\beqa
&&K=k\tlam=\sqrt{1-\frac43\frac{V''(\pii)}{V(\pii)}},
\label{k}\\
&&F(a)=\sqrt{1+(\Omp^{-1}-1)a^{-3}}.
\label{fa}
\eeqa
$t(a)$ can be derived from Eq. (\ref{scale}) so that 
\beqa
{kt(a)}=K\sinh^{-1}\sqrt{\frac{a^3\Omp}{1-\Omp}}=
K\ln \left[\sqrt{\frac{a^3\Omp}{1-\Omp}}(1+F(a))\right].
\label{eka}
\eeqa
Then Eq. (\ref{eos2}) can be written as
\beqa
1+w(a)=(1+w_0)a^{3(K-1)}\left(\frac{(K-F(a))(F(a)+1)^K+(K+F(a))(F(a)-1)^K}
{(K-\Omp^{-1/2})(\Omp^{-1/2}+1)^K+(K+\Omp^{-1/2})(\Omp^{-1/2}-1)^K}\right)^2.
\label{eos3}
\eeqa
Remarkably, the expression Eq. (\ref{eos3}) formally coincides with that 
of \cite{ds} (Eq. (31) in \cite{ds}) 
where the expression is derived for hilltop quintessence. However, the definition of $K$ is 
different: $K=\sqrt{1-(4/3)V''/V}$ in \cite{ds} where $V$ and $V''$ are evaluated at 
the {\it maximum} of $V$, while our $K$ (Eq. (\ref{k})) is evaluated at the {\it initial} 
value $\pii$. 
Moreover, we derive this expression for a general Taylor 
expanded potential around $\pii$. Hence Eq. (\ref{eos3}) is not limited to hilltop 
quintessence but can be applied to a much wider class of slow-roll quintessence. 
It should be noted, however, that since $w(a)$ is an increasing function of $a$, which is 
evident from the derivation, Eq.(\ref{eos3}) can be applied only to thawing model: model with 
growing $w$. 

We also note that we can consider the case of $K^2<0$ (or $V''>(3/4)V$) by the replacement: $K=iK'$ 
where $K'=\sqrt{(4/3)(V''/V)-1}$. Using Eq. (\ref{eka}), $w(a)$ corresponding to Eq. (\ref{eos2}) is give by 
\beqa
1+w(a)={(1+w_0)}{a^{-3}}\left(\frac{K'\cos \left(K'\sinh^{-1}\sqrt{\frac{a^3\Omp}{1-\Omp}}\right)
-F(a) \sin \left(K' \sinh^{-1}\sqrt{\frac{a^3\Omp}{1-\Omp}}\right) }
{K'\cos\left(K'\sinh^{-1}\sqrt{\frac{\Omp}{1-\Omp}}\right)-
\Omp^{-1/2}\sin\left(K'\sinh^{-1}\sqrt{\frac{\Omp}{1-\Omp}}\right)}\right)^2.
\eeqa

\subsection{Consistency Conditions}

\begin{figure}
\includegraphics[width=12cm]{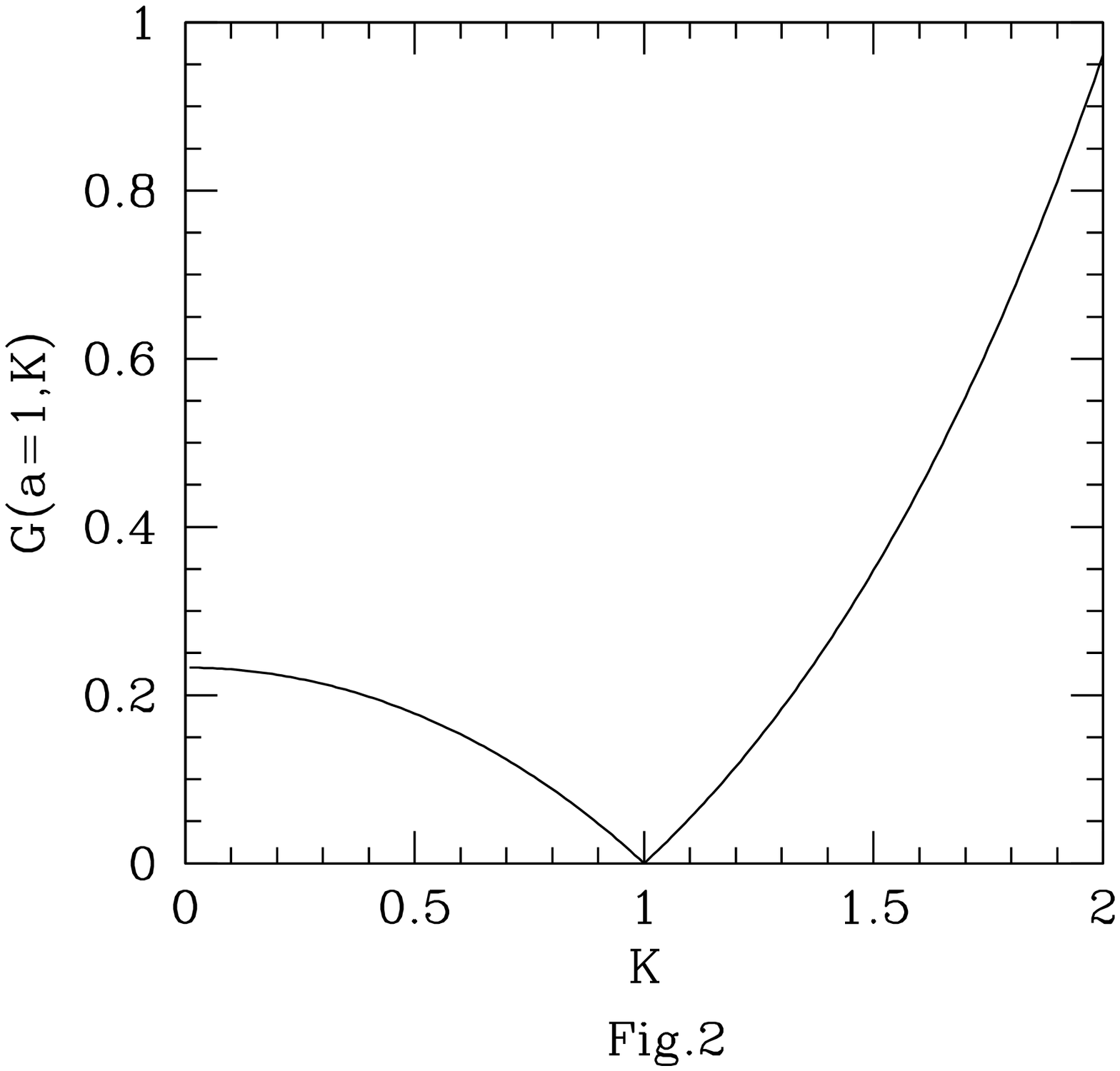}
\caption{ $G(a=1,K)$ as a function of $K$ for $\Omp=0.74$. }
\label{fig2}
\end{figure}

\begin{figure}
\includegraphics[width=13cm]{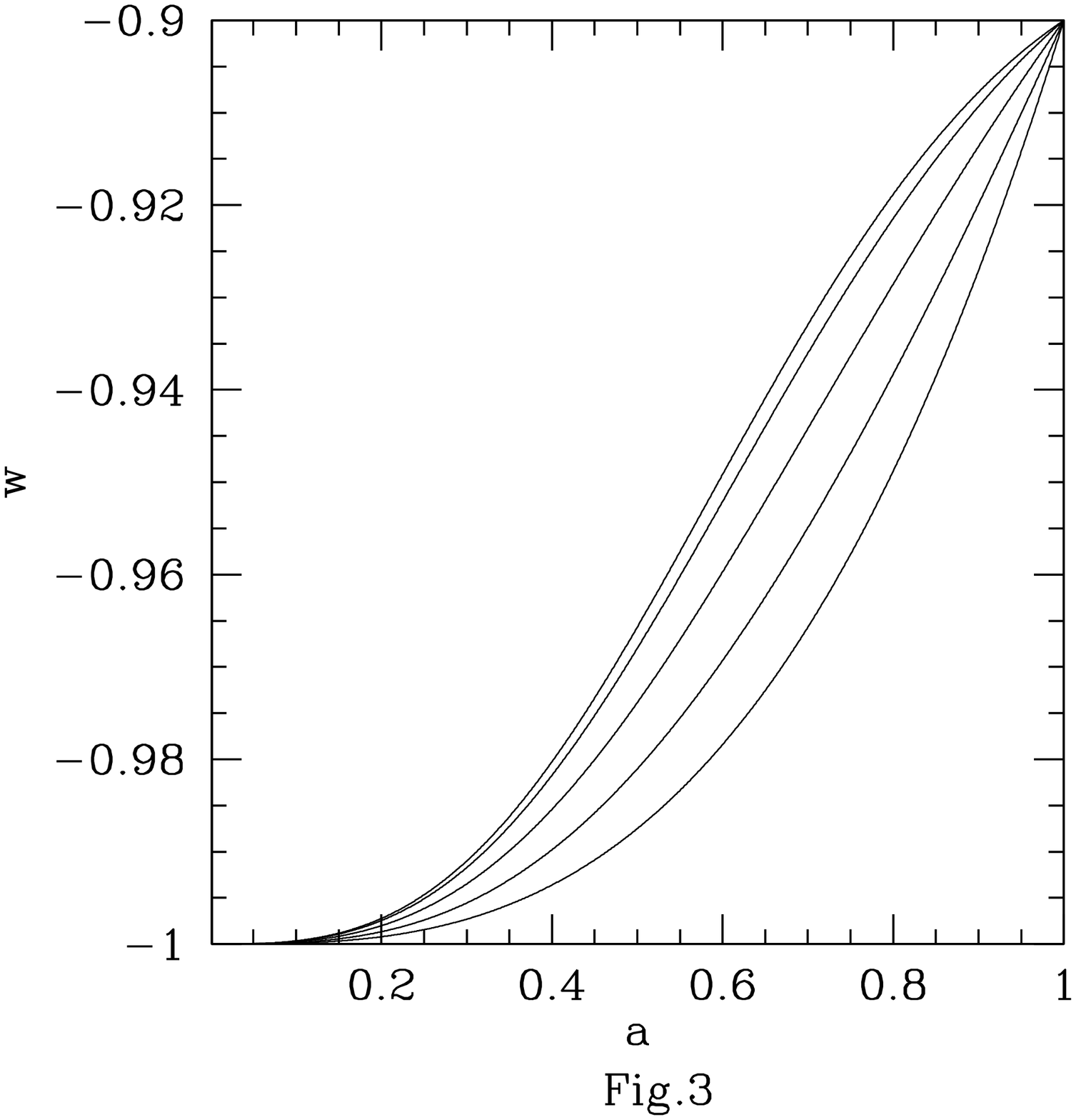}
\caption{$w(a)$ for several $K$ with $\Omp=0.74$. $K=0.01,0.5,1,1.5,2$ from left to right. }
\label{fig3}
\end{figure}

Now, we check the consistency of the expansion Eq. (\ref{exp}) and the assumption $|1+w|\ll 1$.  
The expansion Eq. (\ref{exp}) is consistent if 
$V(\pii)> |V'(\pii)(\p-\pii)|> |(1/2)V''(\pii)(\p-\pii)^2|$. 
Using Eq. (\ref{phi}), this is satisfied if
\beqa
\Gamma:=\Bigl{|}{V(\pii)V''(\pii)\over V'(\pii)^2}\Bigr{|}={\cal O}(1),\label{slow1}\\
G(a,K):=\Bigl{|} \frac{\sinh(kt)}{K\sinh(t/\tlam)} -1 \Bigr{|}< 1 ,
\label{limitk}
\eeqa
where we have introduced $\Gamma$ first introduced (without the absolute value) in \cite{swz}. 
Note that $G(a,K)$ is an increasing function of $a$. Using Eq. (\ref{eka}), $G(a=1,K)$ is given by
\beqa
G(a=1,K)=\Bigl{|}\frac{1}{2K}\left({\Omp\over 1-\Omp}\right)^{(K-1)/2}
\left((\Omp^{-1/2}+1)^K-(\Omp^{-1/2}-1)^K\right)-1\Bigr{|}
\eeqa
In Fig. 2, $G(a=1,K)$ is plotted for $\Omp=0.74$. It is  smaller than unity and 
the inequality Eq. (\ref{limitk}) is satisfied if $|K-1|< 1$. 
In terms of the potential, this implies
\beqa
\Bigl{|}\frac{V''(\pii)}{V(\pii)}\Bigr{|}< 1.
\label{slow2}
\eeqa

Since the epoch of the initial conditions is arbitrary as long as $a_i\ll 1$, 
$\pii$ in Eq. (\ref{slow1}), Eq. (\ref{slow2}) may be replaced with $\p$. 
Then we finally obtain the consistency conditions of slow-roll thawing quintessence: 
\beqa
\Gamma=\Bigl{|}{V(\p)V''(\p)\over V'(\p)^2}\Bigr{|}={\cal O}(1),\label{cond1}\\
\Bigl{|}\frac{V''(\p)}{V(\p)}\Bigr{|}< 1. \label{cond2}
\eeqa
Note that from $(V'/V)^2=\Gamma^{-1} |V''/V|$, these two conditions imply $(V'/V)^2<1$. 
Since $3H^2\gg V$ during matter/radiation epoch, $|V''/V|<1$ and $(V'/V)^2<1$ are consistent 
with the slow-roll conditions, Eq. (\ref{slow:cond:1}) and Eq. (\ref{slow:cond:2}), which 
ensures $|1+w|\ll 1$.

\subsection{$K=1$ Case}

Lastly, for completeness, we consider the case of $K=1$ (or $V''(\pii)=0$) \cite{ss}. 
In this case, the general solution of Eq. (\ref{eomu3}) with $\p=\pii$ at $t=0$ is
\beqa
\p(t)=\pii+\frac{2V'(\pii)}{3V(\pii)}\left(1-\frac{kt}{\tanh (kt)}\right).
\eeqa
The equation of state is
\beqa
1+w&=&\frac13\left(\frac{V'(\pii)}{V(\pii)}\right)^2\left(\frac{\sinh (kt)\cosh (kt)-kt}{\sinh^2(kt)}\right)^2\nonumber\\
&=&(1+w_0)\left(\frac{
F(a)-\frac{1-\Omp}{\Omp a^3}\ln \left[\sqrt{{\Omp a^3\over 1-\Omp}}(1+F(a))\right]}
{\Omp^{-1/2}-\frac{1-\Omp}{\Omp}\ln \left[\frac{1+\Omp^{1/2}}{\sqrt{1-\Omp}}\right]}\right)^2.
\label{eos31}
\eeqa
Therefore, the Taylor expansion is consistent if
\beqa
\left(\frac{V'(\pii)}{V(\pii)}\right)^2< 1, 
\eeqa
which is again consistent with the slow-roll condition, Eq. (\ref{slow:cond:1}), and hence ensures 
$|1+w|\ll 1$. 
Eq. (\ref{eos31}) is identical to the expression derived for a linear potential in \cite{ss}. 
It is easily found that Eq. (\ref{eos3}) reduces to Eq. (\ref{eos31}) in the limit 
$K\rightarrow 1$, confirming the result by \cite{ds}.

In Fig. 3, $w(a)$ given by Eq. (\ref{eos3}) or by Eq. (\ref{eos31}) is shown for several $K$.

\section{Comparison}

\subsection{Comparison with Numerical Solutions}

\begin{figure}
\begin{minipage}{0.31\linewidth}
\includegraphics[width=\linewidth]{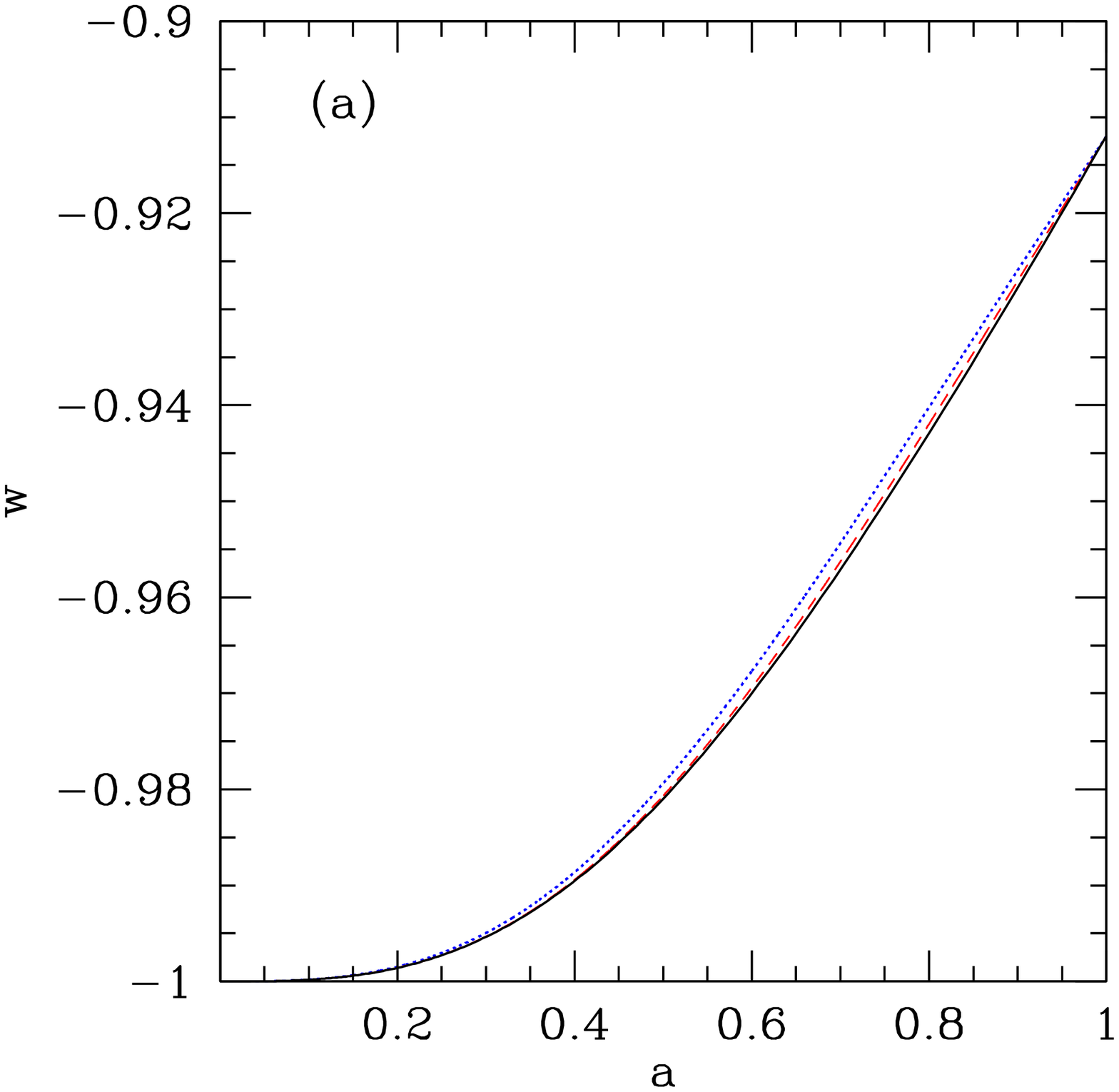}
\end{minipage}
\begin{minipage}{0.31\linewidth}
\includegraphics[width=\linewidth]{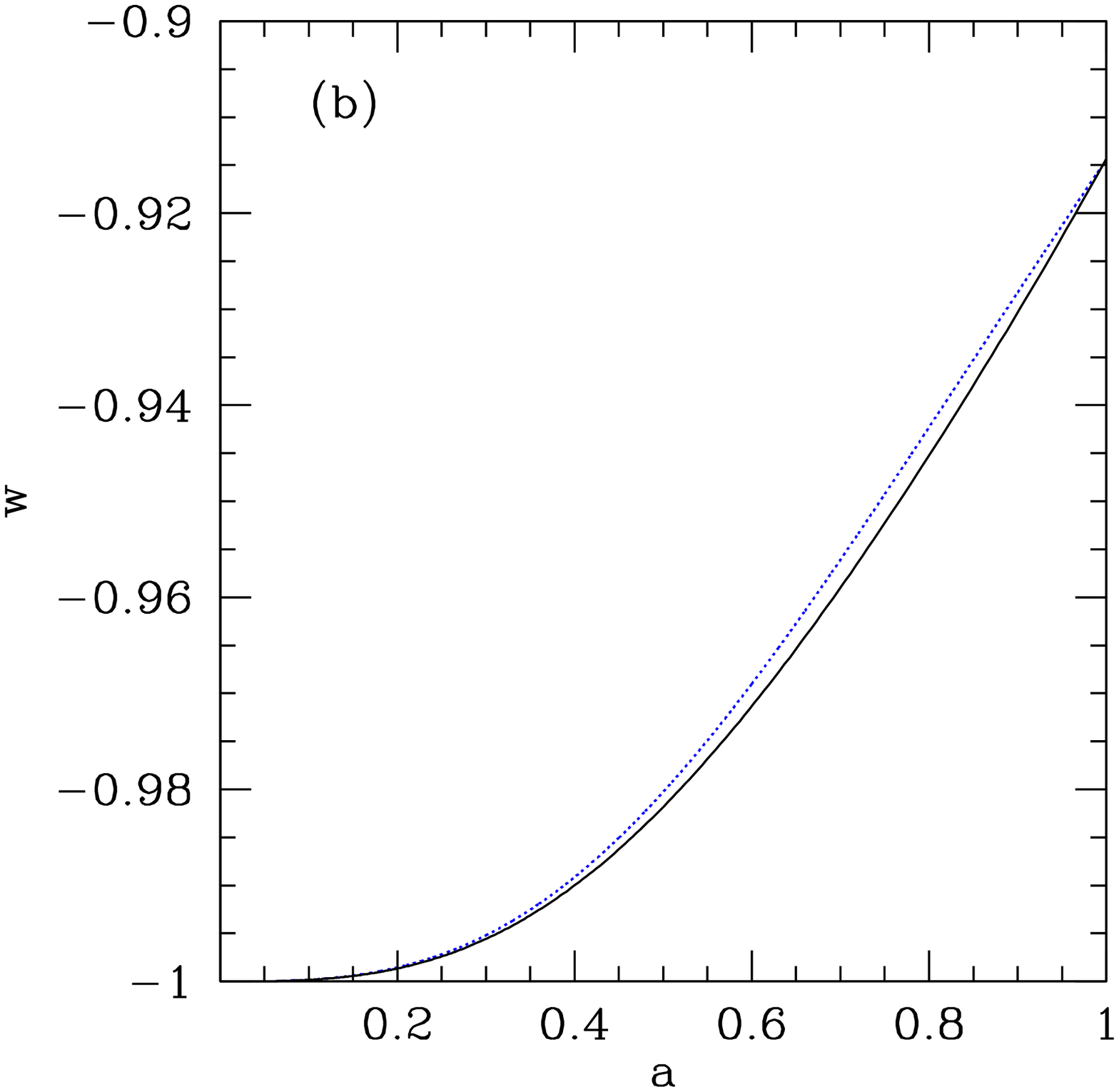}
\end{minipage}
\begin{minipage}{0.31\linewidth}
\includegraphics[width=\linewidth]{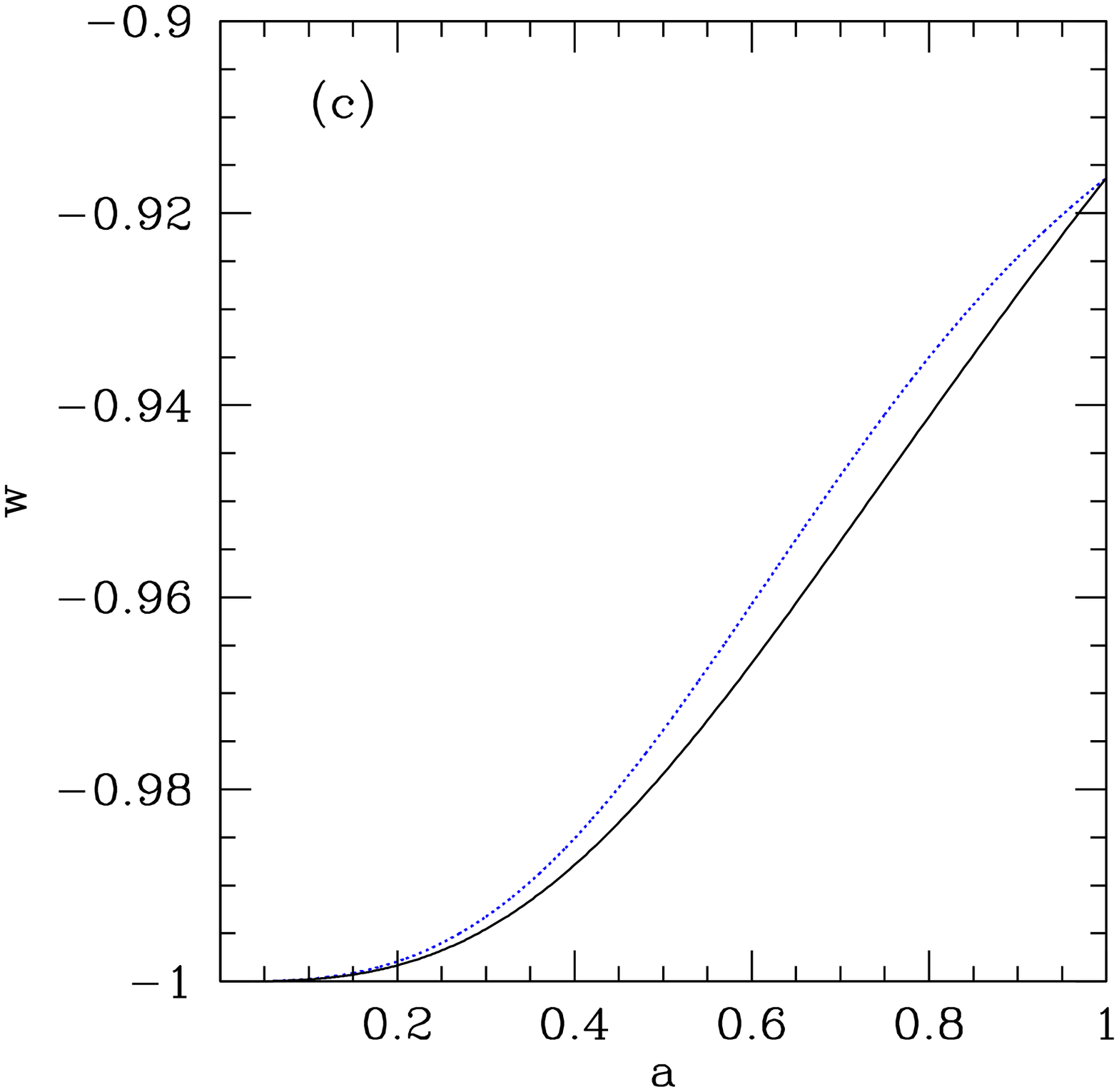}
\end{minipage}
\caption{$w(a)$ for (a) the axion-like potential, $V=M^4[1-\cos\p]$, 
for (b) the logarithmic potential, $V=M^4\log \p$, and for (c) the quadratic potential, $V=m^2\p^2/2$. 
The solid (black) curve is the numerical solution, the dotted (blue) curve gives 
our approximation, and the dashed (red) curve gives the approximation in \cite{ds}.  }
\label{fig4}
\end{figure}

We compare the slow-roll prediction of $w(a)$ (Eq. (\ref{eos3})) with numerical solutions 
for several models and evaluate the accuracy of Eq. (\ref{eos3}). 
We consider the following three examples: 

\paragraph*{(a)} the pseudo Nambu-Goldstone boson (axion-like) model \cite{pngb}:
\beqa
V(\p)=M^4\left(1-\cos(\p/f)\right),
\eeqa
\paragraph*{(b)} logarithmic potential \cite{log}: 
\beqa
V(\p)=M^4\log(\p),
\eeqa 
\paragraph*{(c)} quadratic potential \cite{quad}:
\beqa
V(\p)={1\over 2}m^2\p^2,
\eeqa
where $M$, $f$ and $m$ are constants. The first example is considered 
in \cite{ds}, and the second example corresponds to the potential without maximum/minimum, and 
the third example corresponds to the concave potential $V''>0$ so that $K<1$. 
We fix $\Omp=0.74$ and take $f=1$ in the reduced Planck units and choose $\p_i$ so 
that $w_0\simeq -0.9$. The results are shown in Fig. \ref{fig4}. 

For all cases, we find fairly good agreement with the numerical solutions: 
For case (a), the relative error (the difference between our approximation and 
the numerical solution), 
$|\delta w/w|$, is 
less than 0.3\% while it is less than 0.1\% for the approximation by \cite{ds}. 
For case (b), $|\delta w/w|\siml 0.3\%$ and for case (c) it is less than 0.7\%. 
Note that the potential does not have a maximum for the latter two cases and the approximation of 
\cite{ds} is no longer available. To check the slow-roll conditions 
Eq. (\ref{cond1}) and Eq. (\ref{cond2}), we compute $\Gamma,  V''/V$ at $\pii$: 
$\Gamma=1.33,0.77,0.50; |V''/V|=0.31,0.28,0.22$, 
respectively.\footnote{It is to be noted that the slow-roll conditions Eq. (\ref{cond1}) and 
Eq. (\ref{cond2}) are 
required for the consistency of the solution Eq. (\ref{eos3}) and define the range of validity of the solution; 
otherwise the expansion of the potential Eq. (\ref{exp}) and $|1+w|\ll 1$ cannot be trusted. However, 
this does not imply 
that Eq. (\ref{eos3}) cannot be used for $|V''/V|\gg 1$ (or $K\gg 1$), but rather simply that 
we cannot trust such an extrapolation. Interestingly, for axion case (and other hilltop quintessence 
model), it is shown that approximation Eq. (\ref{eos3}) with $K$ evaluated at the maximum 
of the potential agrees excellently with the numerical solution even for $K=4$ \cite{ds}.}

\subsection{Comparison with Other Parametrizations}

Finally we compare our parametrization with other parametrizations of $w(a)$. 

The most frequently used functional form is the linear approximation of $w(a)$ at $a=1$, 
the so-called Chevallier-Polarski-Linder parameterization, $w_{\rm linear}(a)$, 
Eq. (\ref{linder}) \cite{cpl}. 


Another parametrization of $w(a)$ closely related our approach is that by Crittenden et al.\cite{cmp}. 
Instead of expanding the potential, they expanded the slow-roll parameter around $\p_0$ 
in linear order: 
\beqa
\ka(\p)=\frac{V'}{(1+\beta)V}=\ka_0+\ka_1(\p-\p_0).
\eeqa
Resulting $w$, denoted as $w_{\rm cmp}(a)$, is written as \cite{cmp}
\beqa
1+w_{\rm cmp}(a)&=&\frac13 \ka^2_0\Omp^{-2\ka_1/3}a^{-2\ka_1}F(a)^{-(4\ka_1+6)/3}\nonumber\\
&=&(1+w_0)\Omp^{-(2\ka_1+3)/3}a^{-2\ka_1}F(a)^{-(4\ka_1+6)/3},
\label{cmp:w}
\eeqa
where in the last line we have normalized the equation of state to the present-day value. 

\begin{figure}
\includegraphics[width=13cm]{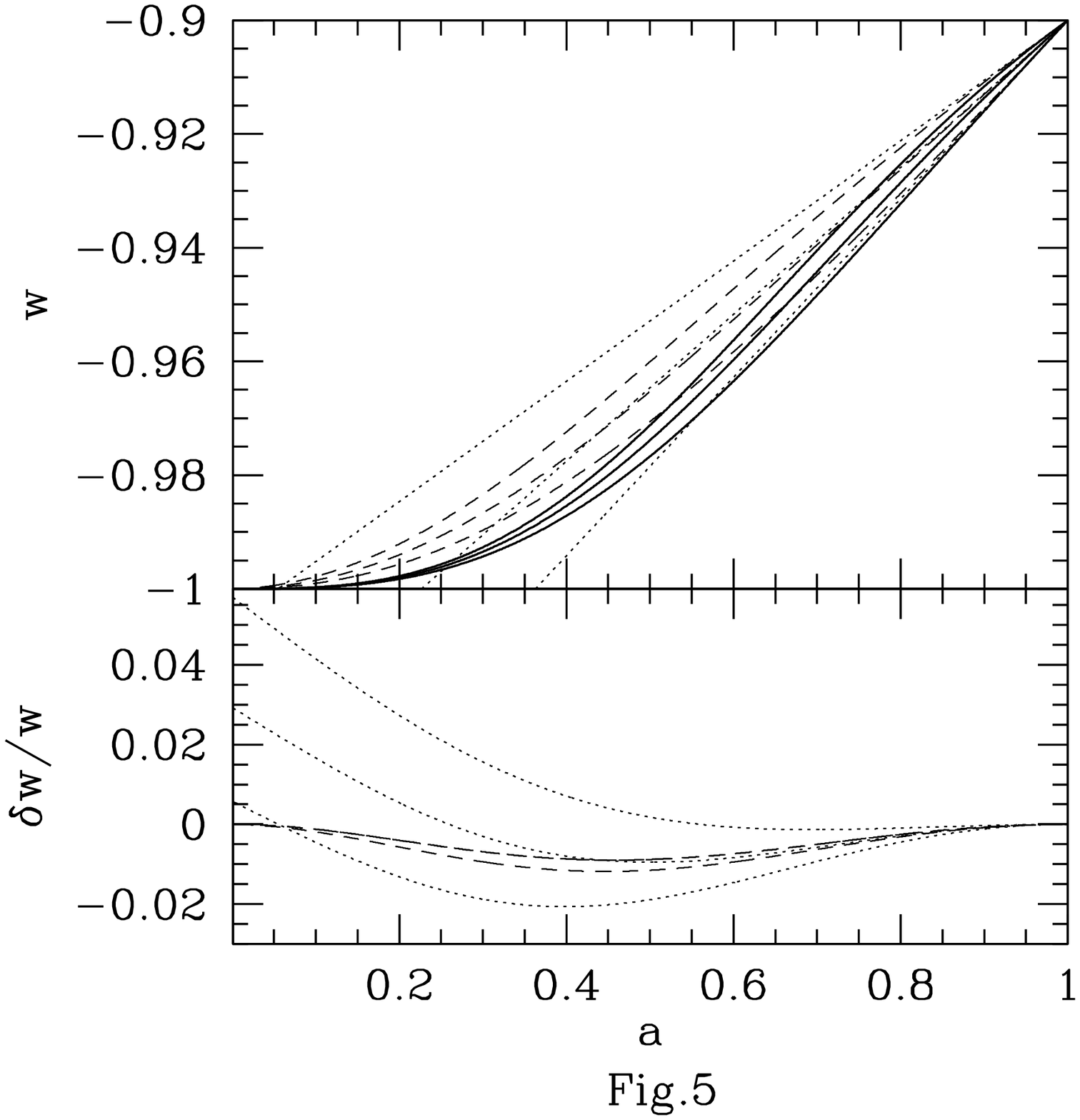}
\caption{Top: Comparison between our parametrization  $w(a)$ (Eq. (\ref{eos3})) (solid curve) and a 
linear parametrization $w_{\rm linear}(a)$ (dotted curve) and 
$w_{\rm cmp}(a)$ \cite{cmp} (dashed curve) 
for $K=0.8,1,1.2$ from left to right. 
Bottom: the relative error between $w_{\rm linear}(a)$ (dotted) or $w_{\rm cmp}(a)$ (dashed) 
and our parametrization: $\delta w/w=(w_{\rm linear/cmp}-w(a))/w(a)$. $K=0.8,1,1.2$ from 
bottom to top. }
\label{fig7}
\end{figure}

In the upper panel of Fig. \ref{fig7}, three parametrizations of the equation of state normalized to 
the present-day value (and 
the first derivative) are shown.  In the lower panel, the relative error is shown. 
It can be seen that the difference between the linear parametrization $w_{\rm linear}(a)$  
and ours is less 2\% for $a\simg 0.5$ 
(or $z\siml 1$), but it can be as large as 6\% for smaller $a$.  On the other hand, 
the difference between $w_{\rm cmp}(a)$ and ours is at most less than 1\%.   
Hence, as far as the goodness of fit is concerned, there is no difference 
between them. However, while for $w_{\rm cmp}(a)$ $\ka_1$ is related (roughly) to the first 
and the second derivative of the potential, for our 
parametrization Eq. (\ref{eos3})  $K$ is directly related to the curvature of the potential. 

It should be stressed that our $w(a)$ Eq. (\ref{eos3}) is {\it not} a fitting function particularly 
designed to match the numerical solutions, but rather a function {\it derived} by solving the equation 
of motion of $\p$. 
Our results demonstrate that only a slight change of the definition 
of $K$ in \cite{ds} as in Eq. (\ref{k}) greatly extends the applicability of 
Eq. (\ref{eos3}). Eq. (\ref{eos3}) can be used not only for hilltop quintessence but also 
for other quintessence with a more general potential without maximum. 
Conversely, we propose that the parametrization of the equation of state of the form 
Eq. (\ref{eos3}) with two free parameters $(w_0,K)$ may be used to fit the cosmological data. 
It fits better than the commonly used linear equation of state, 
$w_{\rm linear}(a)=w_0+w_a(1-a)$, and 
more importantly the meaning of the parameter $K$ is clear: the curvature of the potential. 

\section{Summary}

We have derived slow-roll conditions for thawing quintessence models, Eq. (\ref{slow:cond:1}) and 
Eq. (\ref{slow:cond:2}). 
We have also solved the equation of motion of the slow-roll thawing quintessence and 
obtained the equation of state as a function of the scale factor $w(a)$, Eq. (\ref{eos3}), which 
involves only two parameters. We have derived the consistency conditions of the approximation, 
Eq. (\ref{cond1}) and Eq. (\ref{cond2}),  which 
are consistent with the slow-roll conditions. 
We have found that  only a slight change 
of the definition of $K$ in \cite{ds} greatly extends the applicability of their $w(a)$. 
We have shown that our $w(a)$ agrees fairly well with the 
numerical solutions for several thawing models and found that our $w(a)$ is in general not fit by 
a linear evolution in $a$ as emphasized by \cite{ds}.  

It would be desirable to have useful approximation of $w(a)$ for freezing quintessence models 
and to obtain the unified expression for $w(a)$. 
However, to do so, the different approach is needed,  
since the equation of state can be significantly different from $-1$ during matter/radiation epoch. 
We have derived slow-roll conditions for freezing quintessence models, Eq. (\ref{slow:cond:1}) and 
Eq. (\ref{etasol2}). 

It would also be interesting to extend the slow-roll conditions to quintessence with non-minimal 
coupling with gravity (extended quintessence) 
\cite{chiba} by extending the results for non-minimally coupled inflaton(s) \cite{cy}, 
which could provide conditions for "tracking without tracking" to solve  
the coincidence problem dynamically \cite{chiba2}.

\section*{Acknowledgments}
The author would like to thank Robert Scherrer for useful correspondence. 
This work was supported in part by Grant-in-Aid for Scientific Research
from JSPS (No.\,17204018, No.\,20540280)
and from MEXT (No.\,20040006) and in part by Nihon University. 
Numerical computations were performed at YITP in Kyoto University.

\appendix
\section{Slow-Roll Conditions for Freezing Quintessence}
\label{app1}

\begin{figure}
\includegraphics[width=13cm]{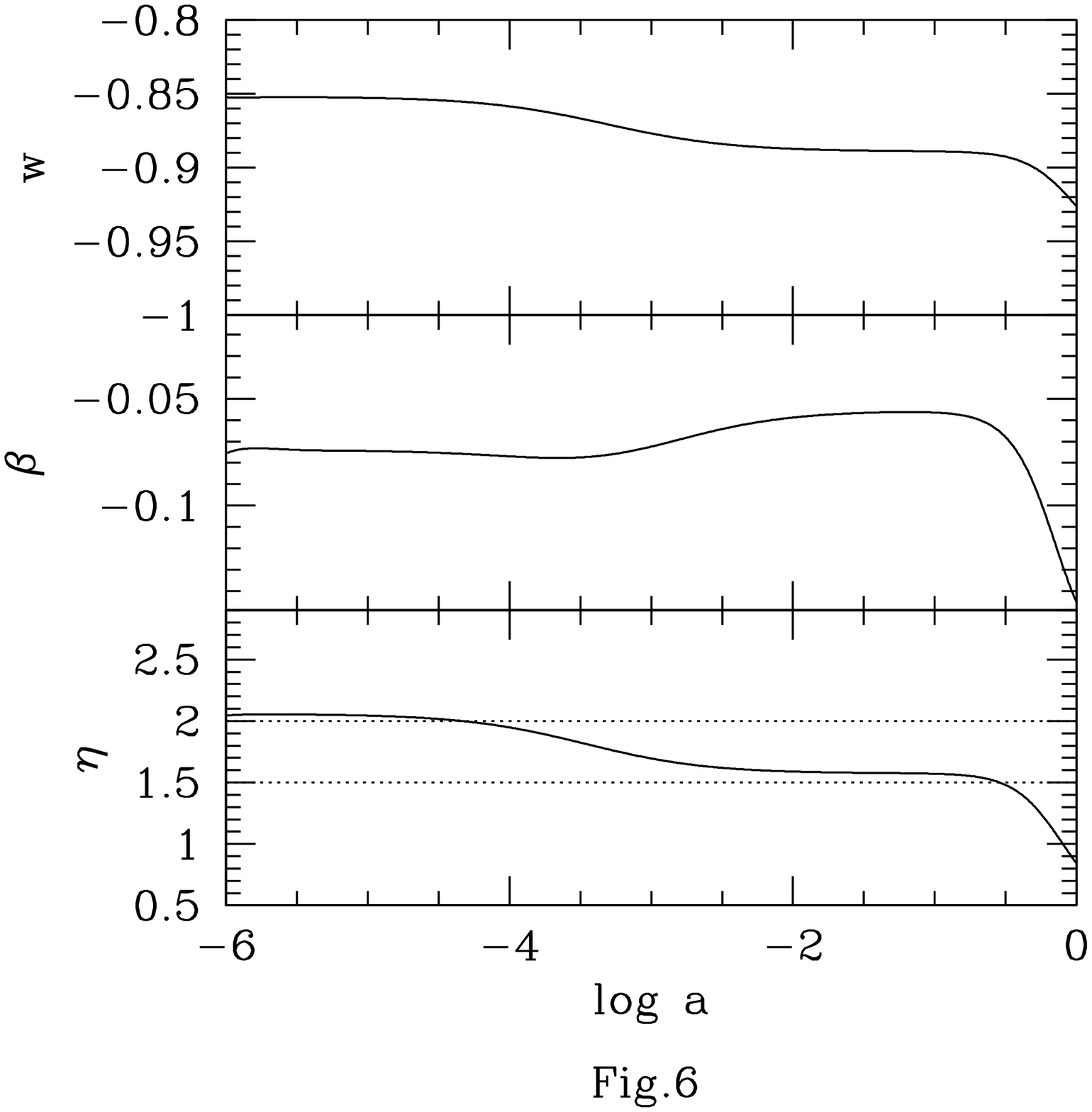}
\caption{ $w, \beta$ and $\eta$ as a function of $a$ for a freezing quintessence model 
with the potential $V=M^4\p^{-1/4}$. The dotted lines are $\eta=2, 3/2$, respectively. }
\label{fig8}
\end{figure}

We derive the slow-roll conditions for freezing quintessence during the matter/radiation epoch. 
For slowly rolling freezing models, although the kinetic energy density is smaller than 
the potential Eq. (\ref{slowroll1}), 
$w$ is not so close to $-1$. Then, compared with slowly rolling thawing models, 
the Hubble friction is effective and so in Eq. (\ref{eomp}) 
\beqa
|\beta|=\left|\frac{\ddot\p}{3H\dot\p}\right|\ll 1,
\label{slowroll2}
\eeqa
 and the usual slow-roll equation of motion is obtained: 
\beqa
3H\dot\p+V'= 0.
\label{pdot2}
\eeqa  
Using Eq. (\ref{pdot2}), the slow-roll condition Eq. (\ref{slowroll1}) becomes again 
$\epsilon\ll 1$, (Eq. (\ref{slow:cond:1})). The time derivative of Eq. (\ref{pdot2}) gives
\beqa
\beta=\frac{\ddot\p}{3H\dot\p}=-\frac{V''}{9H^2}+\frac{1+w_B}{2}.
\eeqa
Therefore, the assumption Eq. (\ref{slowroll2}) is consistent if 
\beqa
\eta=\frac{V''}{3H^2}=\frac32 (1+w_B),
\label{etasol2}
\eeqa
which coincides with Eq. (\ref{etasol}). Eq. (\ref{slow:cond:1}) and Eq. (\ref{etasol2}) 
constitute the slow-roll conditions for freezing models. In Fig. \ref{fig8}, the evolution of $w, \beta$ 
and $\eta$ for a freezing model $(V=M^4\p^{-1/4})$ is shown. Although $\eta$ deviates from 
Eq. (\ref{etasol2}) to the extent that $\beta$ deviates from zero (or $w$ deviates from $-1$), 
we find reasonable agreement 
with Eq. (\ref{etasol2}).

\section{Reduction of Eq. (\ref{phi:general}) to Eq. (\ref{phi})}
\label{app2}
The general solution of Eq. (\ref{eomu3}) with $\p=\pii$ and $\dot\p=\dot\pii$ at $t=t_i$ 
is given by Eq. (\ref{phi:general}). Using Eq. (\ref{scale}) and Eq. (\ref{eka}), 
we obtain
\beqa
e^{kt_i}=\left(\frac{\Omp}{1-\Omp}\right)^{K/2}a_i^{3K/2}(F(a_i)+1)^K.
\eeqa
For $a_i\ll 1$, since $F(a_i)\simeq \sqrt{\Omp^{-1}-1}a_i^{-3/2}$ from Eq. (\ref{fa}),
we obtain
\beqa
\sinh(kt_i)\simeq K\sqrt{\frac{\Omp a_i^3}{1-\Omp}},~~~~~\cosh(kt_i)\simeq 1+{\cal O}(a_i^3).
\eeqa
We also note that
\beqa
\sinh(t_i/\tlam)=\sqrt{\frac{\Omp a_i^3}{1-\Omp}},~~~~~\cosh(t_i/\tlam)=F(a_i)\sinh(t_i/\tlam)
\eeqa
Hence, in the limit of $a_i\ll 1$, only the term proportional to $\sinh(kt)$ survives in 
Eq. (\ref{phi:general}) (up 
to ${\cal O}(a_i^{3/2})$). 
Thus, Eq. (\ref{phi:general}) is reduced to Eq. (\ref{phi}). 



\end{document}